\documentclass[journal]{rmaa}

\usepackage{paralist}

\usepackage{psfrag,color}

\usepackage[latin1]{inputenc}

\usepackage{amsmath}
\usepackage{adjustbox}
\usepackage{multirow}

\title{Assessing the Hierarchical Hamiltonian Splitting Integrator for Collisionless \emph{N}-body Simulations} 

\author{
  G. Aguilar-Arg\"uello,\altaffilmark{1} 
  O. Valenzuela,\altaffilmark{1}
  H. Vel\'azquez,\altaffilmark{1}
  J.C. Clemente,\altaffilmark{1}
  and J. A. Trelles\altaffilmark{1}}

\altaffiltext{1}{Instituto de Astronom\'ia, Universidad Nacional Aut\'onoma de M\'exico, A.P. 70--264, 04510, M\'exico, CDMX, M\'exico}

\shortauthor{Aguilar-Arg\"uello}
\shorttitle{Hierarchical Hamiltonian Splitting}

\listofauthors{G. Aguilar-Arg\"uello }
\indexauthor{Aguilar-Arg\"uello, G.}
\indexauthor{Valenzuela, O.}

\abstract{The large dynamic range in some astrophysical \emph{N}-body problems led to the use of adaptive multi-time-steps; however, the search for optimal strategies is still challenging. We numerically quantify the performance of the hierarchical Hamiltonian Splitting (HHS) integrator for collisionless simulations using a direct summation code. We compare HHS with the constant global time-step leapfrog integrator, and with the adaptive one (AKDK). We find that HHS is approximately reversible, whereas AKDK not. Therefore, it is possible to find a combination of parameters where the energy drift is considerably milder for HHS, resulting in a better performance. We conclude that HHS is an attractive alternative to AKDK, and it is certainly advantageous for direct summation and P3M codes. Also, we find advantages with GADGET4 (Tree/FMM) HHS implementation that are worth exploring further. }

\resumen{El gran rango din\'amico en algunos problemas astrof\'isicos de \emph{N}-cuerpos ha llevado al uso de pasos de tiempo m\'ultiples adaptivos, sin embargo, la b\'usqueda de estrategias \'optimas es a\'un un reto. Estudiamos num\'ericamente el integrador Hierarchical Hamiltonian Splitting (HHS) utilizando un c\'odigo de suma directa y comparamos con el rendimiento de leapfrog con paso global constante y su version multi-paso adaptivo (AKDK). Encontramos que HHS es aproximadamente reversible, mientras que AKDK no. Por lo que es posible encontrar una combinaci\'on de par\'ametros tales que el cambio de energ\'ia es considerablemente menor para HHS, resultando en una mayor eficiencia. Concluimos que HHS es una alternativa competitiva con ventaja para c\'odigos de suma directa y P3M. Tambi\'en, hallamos ventajas para la implementaci\'on de HHS en GADGET4 (\'Arbol/FMM) que merecen ser investigadas m\'as. }

\addkeyword{software: simulations }
\addkeyword{gravitation}
\addkeyword{celestial mechanics}
\addkeyword{methods: numerical}
\addkeyword{galaxies: kinematics and dynamics}
\addkeyword{cosmology: large scale structure}

\begin{document}
% Typeset article header
\maketitle

\section{Introduction} \label{sec:intro}

Historically, fully self-consistent realistic astrophysical \emph{N}-body simulations are a challenging problem \citep{1971Ap&SS..14...20A,1985ApJS...57..241E,2001PhDT........21S,2001NewA....6...79S,2011EPJP..126...55D,2017Rev..Anatoly}. On the purely gravitational case, direct summation \emph{N}-body codes suffer from a bottleneck due to the computational cost of force calculation on a particle-by-particle basis, which scales as the square of particle number ($N^2_p$). For this reason, they are primarily used for simulating dense stellar environments or planetary systems. Such limitations triggered the development of sophisticated approximated hybrid collisionless methods, like the TreePM (Tree-Particle Mesh)/P3M (Particle Particle- Particle Mesh)/P3T (Particle Particle- Particle Tree) codes \citep{1995ApJS...98..355X,2000ApJS..128..561B,2002JApA...23..185B,2003ApJS..145....1B} where the short-range component of the force is carried out either by expensive/accurate direct summation or Tree force solvers \citep{1991ApJ...368L..23C,2011PASJ...63..881O,2013hpcn.confE...6H}. Alternatively, AMR (Adaptive Mesh refinement) methods are used to compute the large dynamical range of the gravitational force \citep{1989ApJS...71..407V,1994JCoPh.115..339J,1997ApJS..111...73K,2002A&A...385..337T}, seeking a balance between accuracy and computational efficiency. \emph{N}-body simulations require both, a fast way to calculate the accelerations and an accurate and efficient integration method to evolve particles in time. The second-order leapfrog symplectic integrator \citep{1967PhRv..159...98V} is the most widely used in collisionless \emph{N}-body simulations \citep[e.g.][]{2017Rev..Anatoly,2021arXiv211205165A}. It is strictly symplectic when a global-constant time-step is adopted; however, this is not suitable for addressing problems with a large dynamical range that are currently studied with modern codes. In the quest to improve efficiency, it is necessary to adopt multiple or adaptive time-steps. The general problem of geometric/symplectic (preserving phase space volume), time-symmetric (recover initial conditions after changing $dt$ for -$dt$) and reversible integrators (recover initial conditions after changing the sign of velocities) has been addressed in the field of differential equations for dynamical systems \citep{2002Springer..}. In such work, they point out the differences between adaptive global time-steps or multi-time-steps (several rates of evolution for different parts of the system) and discuss the constraints required for the integration method and the time-step selection function to preserve the mapping properties. In astronomy, the influential study of \citet{1997astro.ph.10043Q} discusses different operator-based leapfrog implementations using time step blocks plus a time-step selection function in the KDK/DKD leapfrog integrator for massive \emph{N}-body simulations. They pointed out that particle migration across time-step blocks may break up the symmetry and sometimes involves backward integration, which can be difficult to reconcile with a dissipative component like gas. Current collisionless simulations codes commonly use the KDK leapfrog implementation with adaptive time-steps \citep[e.g.][]{1997astro.ph.10043Q,2005MNRAS.364.1105S,2011EPJP..126...55D,2017Rev..Anatoly}, which we call hereafter AKDK. Recently, \citet{2017MNRAS.472.1226D} discusses conditions where the \citet{2002Springer..} analysis for time-symmetric integrators can be extended to discrete time-stepping. They conclude that there is not a general solution. Many of the proposed integrators truly preserve the symmetries; however, the specific time-step selection function should also respect symmetrization. In several cases, the computational overhead makes the proposal impractical.

In this paper, we explore and quantify the Hierarchical Hamiltonian Splitting (HHS) strategy proposed by \citet{2012NewA...17..711P}, which is, as leapfrog, a second-order scheme. This integrator was tested for small number of bodies or collisional simulations, delivering good energy conservation and momentum conservation at machine accuracy. However, no analysis of time-symmetry or reversibility was presented. Some modern \emph{N}-body codes like AREPO \citep{2020ApJS..248...32W} and GADGET4 \citep{2021MNRAS.506.2871S} have implemented versions of HHS with some differences with respect the original proposal, although they do not give details of the performance or other properties that allow the comparison with the commonly used integrators. In this work, we extend the discussion of HHS in the context of collisionless \emph{N}-body simulations, by numerically investigating the time-symmetry and velocity reversibility. We, also, test some time-step selection functions to explore the potential advantages. In all cases we compare with the global constant time-step leapfrog integrator and the adaptive one (AKDK) in order to asses the conditions under which HHS is a competitive alternative. As we discussed above, in some modern P3M codes running in hybrid architectures, the most expensive calculation is the short-range direct summation force integration, in some cases processed inside GPUs \citep{2016NewA...42...49H}. Motivated by that we implemented HHS in a direct summation code running in GPUs and complement that with additional tests with the Tree/FMM code GADGET4.

The rest of the paper has been organized as follows: section \ref{Integrators} summarizes the main properties of the integrators used here to carry out their comparison while section \ref{TSelect} introduces the time-step selection functions. Section \ref{acc_Tests} shows accuracy tests performed with emphasis on the cases on an isolated halo and a minor merger. Sections \ref{TS_tests}, \ref{Perfomance} and \ref{Stability} contain the results of these tests taking into account the effect of time-step functions, performance and long-term stability, respectively. In section \ref{Reversibility}, we quantified reversibility and time-symmetry for the different codes. Finally, a discussion and main conclusions are given in section \ref{Conclusions}.

\section{Integrators} \label{Integrators}

We implemented three different integrators in a direct summation code dubbed as $\mathrm{NP_{splitt}}$ (Aguilar-Arg\"uello et al. \emph{in prep.}), the leapfrog, Adaptive-KDK (AKDK) and the Hierarchical Hamiltonian Splitting integrators (HHS). Below, we describe each integrator. It is common to express integrators as a composition of operators using the Hamiltonian splitting technique in potential (\emph{Kick}) and kinetic energy (\emph{Drift}), although there are other possibilities \citep{2011PASJ...63..881O}.

\subsection{Leapfrog}
The leapfrog integrator is a second-order widely used integrator. As mentioned previously, this integrator is strictly symplectic only when a global-constant time-step is adopted. Symplectic integrators are designed to numerically preserve the integrals of motion and the phase-space volume of the simulated system.

In the leapfrog method, the evolution of the gravitational system can be written as a sequence of \emph{Kick} (advance of velocities) and \emph{Drift} (advance of positions) operators \citep[e.g.][]{Channell...1993,1997astro.ph.10043Q}, defined by:
\begin{equation}
    \begin{alignedat}{1}
        K(dt): \textbf{v}\,(t_{n}+dt)\, = & \, \textbf{v}\,(t_{n})\,+ \, dt\, \textbf{a}\,(t_{n}) \\
        D(dt): \textbf{x}\,(t_{n}+dt)\, = & \, \textbf{x}\,(t_{n})\,+ \,dt\, \textbf{v}\,(t_{n})
    \end{alignedat}
\label{eq:1}
\end{equation}
where $\textbf{x}$, $\textbf{v}$ and $\textbf{a}$ are the position, velocity and acceleration of a particle, respectively, and $dt$ is the time step. In this paper, we use the operator sequence called KDK leapfrog \citep[also known as \emph{velocity Verlet,}][]{1982JChPh..76..637S}:
\begin{equation}
    KDK: K(dt/2)\,D(dt)\,K(dt/2)
\label{eq:KDK}
\end{equation}
where we considered that the evolution is for one time step, i.e. from $t_{n}$ to $t_{n}+dt$. Through this paper, we will refer to KDK leapfrog with a global-constant time-step as the Leapfrog integrator.

\subsection{AKDK}
Contemporary codes have extensively used KDK (eq. \ref{eq:KDK}) combined with a block time-step scheme \citep{1967NASSP.153..315H,1985MNRAS.217..127S,1986LNP...267..156M,1989ApJS...70..419H,1991ApJ...369..200M}, frequently using rungs which are power of two: $dt_r = dt_0 2^{(-r)}$ and different assigning time-step functions, most frequently an acceleration based one \citep{2005MNRAS.364.1105S}. We will use it as a reference integration scheme, but it should be noted that it is not symplectic \citep{2002Springer..} and that the block-step is a multi time-step scheme.

\subsection{Hierarchical Hamiltonian Splitting}
The hierarchical Hamiltonian Splitting (HHS) method is a second-order integrator that uses individual time steps of the particles \citep{2012NewA...17..711P} through recursively splitting the Hamiltonian. It accurately preserves linear and angular momentum and has a good energy conservation.

This integrator consists of adaptively and recursively splitting the Hamiltonian as a function of the current time step, $dt$, so that the so called \emph{Slow} system (hereinafter \emph{S}) contains all the particles with a time step larger than $dt$, and the so called \emph{Fast} system (hereinafter \emph{F}) contains all the particles with a time step smaller than $dt$. Thus, the splitting is as follows:
\begin{equation}
    \begin{aligned}
        H_{S}= & T_{S}+V_{SS}+V_{SF}\\
        H_{F}= & T_{F}+V_{FF}
    \end{aligned}
\label{eq:8}
\end{equation}
where,
\begin{equation}
    \begin{aligned}
        T_{X}\equiv & \sum_{i\in X}\frac{p_{i}^{2}}{2m_{i}}\\
        V_{XX}\equiv & -G\sum_{i\in X}\sum_{j\in X,\, j>i} \frac{m_{i}m_{j}}{\left|r_{i}-r_{j}\right|}\\
        V_{XY}\equiv& -G\sum_{i\in X}\sum_{j\in Y} \frac{m_{i}m_{j}}{\left|r_{i}-r_{j}\right|} 
    \end{aligned}
\label{eq:9}
\end{equation}
are, respectively, the kinetic and potential energies, and $V_{SF}$ is the potential energy of the interactions between \emph{S} and \emph{F} particles. The previous splitting scheme is known as HOLD \citep[since it ``holds'' $V_{SF}$ for evaluation at the slow time-step,][]{2012NewA...17..711P}.

The \emph{S} system is solved using the DKD scheme \citep[also known as \emph{position Verlet,}][]{1990JChPh..93.1287T}, which consists of \emph{drifts} of the particles in this system (due to $T_S$) and \emph{kicks} on the particles of both systems (due to $V_{SS}+V_{SF}$). For the \emph{F} system, the same procedure as for the original system is applied but using a halved time-step. Hence, the splitting is applied recursively to the \emph{F} system with time-step $dt/2^{r}$. The recursion ends when the system \emph{F} (of the rung $r$) has no particles. At the end of the current integration step, the new time-step of a particle is calculated. In this scheme, a particle can change its time step to higher (lower) value if its current integration time is synchronized with the higher (lower) rung.

It is well known that the \emph{Kick} and \emph{Drift} operators are symplectic, however, using multiple or adaptive time-steps may not preserve such properties in a general way \citep{2002Springer..}. Therefore we need to investigate the behaviour of HHS.

\section{Time-step Selection Function} \label{TSelect}

Besides the formulation of integrators with individual time steps based on symmetric operators, the choice for each particle time step is made through the so called time-step selection function. There is not a unique choice, arguably the most commonly used time-step function in contemporary collisionless \emph{N}-body codes \citep[e.g. GADGET,][]{2005MNRAS.364.1105S} is based on the acceleration as:
\begin{equation}
    \tau_{i} = \eta_\mathrm{accel}\sqrt{\frac{\epsilon}{a_i}}
\label{eq:standard}
\end{equation}
where $a_i$ is the acceleration acting on the particle $i$, giving the code the possibility of adapting to high/low accelerations and $\epsilon$ is the force plummer softening\footnote{We will adopt the softening as twice the average interparticle distance at the minimum radius where the density profile is not dominated by Poisson fluctuations, as it is usually adopted in collisionless simulations.}. Improvements have been recently discussed by using a dynamical time proxy \citep{2007MNRAS.376..273Z} and a tidal force time scale \citep{2011EPJP..126...55D,2020MNRAS.495.4306G}, establishing a balance between short and long time steps, which may translate into higher efficiency. Extensive comparisons of AKDK with both choices have been discussed in \citet{2007MNRAS.376..273Z} and \citet{2020MNRAS.495.4306G}.

In our study, in an attempt to preserve the energy stability of the HHS integrator while allowing adaptive multi time-steps, and following \citet{2012NewA...17..711P}, we use an approximated time-symmetrized time-step extrapolation criterion for each particle. To obtain such time-step criterion, we start from the implicit criterion \citep{1995ApJ...443L..93H}:
\begin{equation}
    \tau_\mathrm{sym}=\frac{1}{2}\left[\tau\left(t\right)+\tau\left(t+\tau_\mathrm{sym}\right)\right]
    \label{eq:imp_sym}
\end{equation}
where $\tau$ is a time-step function (non-symmetrized), and $\tau_\mathrm{sym}$ is the symmetrized time-step function of $\tau$. To a first-order perturbative expansion:
\begin{equation}
    \tau\left(t+\tau_\mathrm{sum}\right) \approx \tau\left(t\right)+\frac{d\tau}{dt}\tau_\mathrm{sym}
\end{equation}
hence, from eq. \ref{eq:imp_sym}:
\begin{equation}
     \tau_\mathrm{sym} \approx \tau(t) + \frac{1}{2}\frac{d\tau}{dt}\tau_\mathrm{sym}
\end{equation}
so that, the time-step we will use is given by \citep{2012NewA...17..711P},
\begin{equation}
    \tau_{i}=\min_{j}\left[\frac{\tau_{ij}}{\left(1-\frac{1}{2}\frac{d\tau_{ij}}{dt}\right)}\right]
    \label{eq:symmHS}
\end{equation}

It is important to state that the minimization indicated above is across the so called \emph{Slow} particles. For a time-step proportional to the inter-particle free-fall time:
\begin{equation}
    \begin{aligned}
        \tau_{ij}= & \, \eta_\mathrm{FF}\sqrt{\frac{r_{ij}^{3}}{G\left(m_{i}+m_{j}\right)}}\\
        \frac{d\tau_{ij}}{dt}= & \, \frac{3\textbf{v}_{ij}\cdot\textbf{r}_{ij}}{2r_{ij}^{2}}\tau_{ij}
    \end{aligned}
    \label{eq:freefall}
\end{equation}

The former option is a two-body-based proxy for the dynamical-time-motivated step function suggested by \citet{2007MNRAS.376..273Z}. 

For completeness with \citet{2012NewA...17..711P}, for a time-step proportional to the inter-particle fly-by time (typically used in collisional problems):
\begin{equation}
    \begin{aligned}
        \tau_{ij}= & \, \eta_\mathrm{FB}\frac{r_{ij}}{v_{ij}}\\
        \frac{d\tau_{ij}}{dt}= & \,  \frac{\textbf{v}_{ij}\cdot\textbf{r}_{ij}}{r_{ij}^{2}}\tau_{ij}\left(1+\frac{G\left(m_{i}+m_{j}\right)}{v_{ij}^{2}r_{ij}}\right)
    \end{aligned}
\label{eq:flyby}
\end{equation}

We will quantify the efficiency of such time-step functions, however, the high acceleration derivatives in the case of collisional problems may require going beyond the first order in the perturbative expansion.

Across the paper, we will mostly use the approximated symmetric free-fall time-step for HHS (defined by eqs. \ref{eq:symmHS} and \ref{eq:freefall}), and only in a few tests we will use the minimum of this and the approximated symmetric fly-by time-step (eqs. \ref{eq:symmHS} and \ref{eq:flyby}). In section \ref{TS_tests}, we will present a comparison with the GADGET4 implementation of HHS \citep{2021MNRAS.506.2871S}, this code uses a time-step function similar to eq. \ref{eq:standard} but the accuracy parameter, $\eta_\mathrm{accel}$, is included inside the square root. For AKDK, we will use the standard time-step function given by eq. \ref{eq:standard}.

Table \ref{tab:integradores} summarizes the combinations of integrators and time-step selection functions used through this work.

\begin{table*}
    \caption{Combinations of integrators and time-step functions used across this paper.}\label{tab:integradores}
    \begin{adjustbox}{width=1.0\textwidth,center}
        \begin{tabular}{ c c c c c}\toprule
            Integrator & Integration & Time-step & Time-step & \multirow{2}{4em}{Reference} \\
            name & scheme & scheme & selection function &  \\  \midrule
            Leapfrog & KDK & Global-constant &  & \citet{1967PhRv..159...98V} \\
            \multirow{2}{2em}{AKDK} & \multirow{2}{2em}{KDK} & \multirow{2}{2em}{BLOCK} &  \multirow{2}{2em}{eq.\,\,\ref{eq:standard}} & e.g. \citet{1997astro.ph.10043Q,2005MNRAS.364.1105S} \\
             &  &  &  & \citet{2011EPJP..126...55D,2017Rev..Anatoly} \\
            HHS & HHS & HOLD & eqs. \ref{eq:symmHS}, \ref{eq:freefall} & \citet{2012NewA...17..711P} \\
            \multirow{2}{2em}{nsHHS} & \multirow{2}{2em}{HHS} & \multirow{2}{2em}{HOLD} & eqs. \ref{eq:symmHS}, \ref{eq:freefall} & \multirow{2}{4em}{This\,\,work} \\
             &  &  & (without $d\tau_{ij}/{dt}$) & \\
            sAKDK & KDK & BLOCK & eqs. \ref{eq:symmHS}, \ref{eq:freefall}  & This work \\ \bottomrule
        \end{tabular}
    \end{adjustbox}
\end{table*}

\section{Accuracy Tests} \label{acc_Tests}

In this section, we present the test results of the HHS algorithm in terms of accuracy by simulating an isolated halo and sinking satellites, and compare them with the global-constant time-step leapfrog and the adaptive one, AKDK. To proceed with the comparison we implemented the different integrators in a direct summation \emph{N}-body code (Aguilar-Arg\"uello et al. \emph{in prep.}). All the experiments were run in a single GPU. As a sanity check, we performed binary system tests (not reported here) and the results are consistent with those reported in other works \citep[e.g.][]{2011EPJP..126...55D,2012NewA...17..711P,2005MNRAS.364.1105S}.

\subsection{Isolated Cuspy Halo} \label{halo_tests}
We adopted as a reference model an equal particle mass, isolated halo following the NFW cuspy density profile predicted by collisionless dark matter cosmological simulations \citep{1997ApJ...490..493N}. The large density range and the corresponding different dynamical times make it a suitable system for an adaptive time-step code. Such tests depend on resolution to actually capture the benefit of individual time-steps as compared with a global-constant time-step scheme. We will use as a reference time scale the dynamical time\footnote{A dynamical time, also called crossing time, is the time taken for a typical particle to cross the system. In this paper, a dynamical time is defined as $t_{\mathrm{dyn}}(r)=\left[ \,r^{3} \,/\, GM(r)\, \right]^{1/2}$, where $r$ and $M(r)$ are the radius and mass, respectively.} at the NFW characteristic radius ($r_s$), since it has been used to study the stability of the halo in other works \citep[e.g.][]{2015MNRAS.447.3693K}.

For the integration of our fiducial model, we adopted $G=1$ (gravitational constant), $M_{\mathrm{vir}}=1$ (virial mass) and $r_{\mathrm{s}}=1$ (scale length, also called characteristic radius), as model units. We will use these model units through the paper.

%------------------------Figure----------------------------------%
\begin{figure*}
     \centering
     \begin{tabular}{cc}
        \includegraphics[width=0.9\textwidth,angle=0]{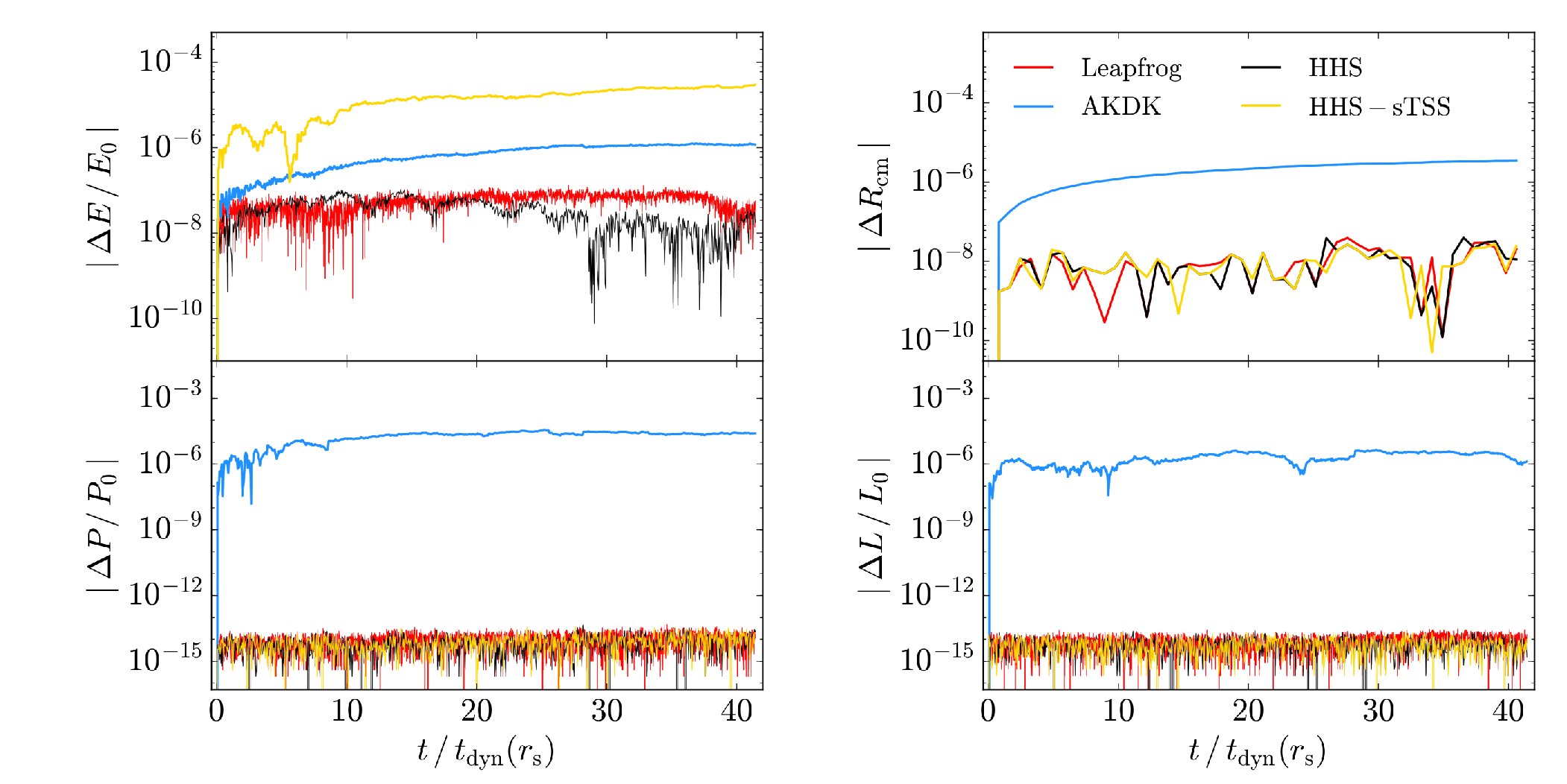}
     \end{tabular}
\caption{Error in conserved quantities for an isolated NFW halo with  $N=10^5$ particles, simulated up to 40 dynamical times (at scale radius, $r_\mathrm{s}$). Shown is the energy error (upper left panel), change in center of mass position (upper right panel), error in linear (lower left panel) and angular momentum (lower right panel) for the three integrators: Leapfrog (red), AKDK (blue) and HHS (black). Also, it is shown the HHS with a delay in the time-step hierarchy update (HHS-sTSS, yellow) version.}
\label{fig:HA_ener_big}
\end{figure*}

To investigate and quantify differences in accuracy and performance between the integrators, first, we followed a fiducial halo sampled with $10^5$ particles for $40$ dynamical times at $r_\mathrm{s}$, $t_\mathrm{dyn}$.

Because our implementations of AKDK and HHS have different time-step function, a meaningful comparison is to assume an energy conservation threshold, which implies using distinct accuracy parameters for both integrators. For the first test, we considered a $10^{-7}$ threshold and accuracy parameters $\eta_\mathrm{FF} =0.003$ and $\eta_\mathrm{accel} = 0.01$ for HHS and AKDK, respectively, both constrained to $6$ time-step rungs. Figure \ref{fig:HA_ener_big} shows the result of these tests. The upper left panel shows the energy error. HHS (black) stays very close to Leapfrog (red) during the first $20\,t_\mathrm{dyn}$, afterwards it shows a small drift. AKDK (blue) drifted almost linearly and after $20\,t_\mathrm{dyn}$ it slightly flattens. Because the main computational overhead of HHS over AKDK comes from building and update the time-step hierarchy in HHS, we decided to explore experiments where we delayed such an update, and denoted by HHS-sTSS. We observed that such an action results in important savings in computational time (yellow line). The energy accuracy test is lower but acceptable for a collisionless simulation and it is faster; in addition linear and angular momentum are preserved to machine precision.

%------------------------Figure----------------------------------%
\begin{figure}
    \centering
    \includegraphics[width=0.49\textwidth,angle=0]{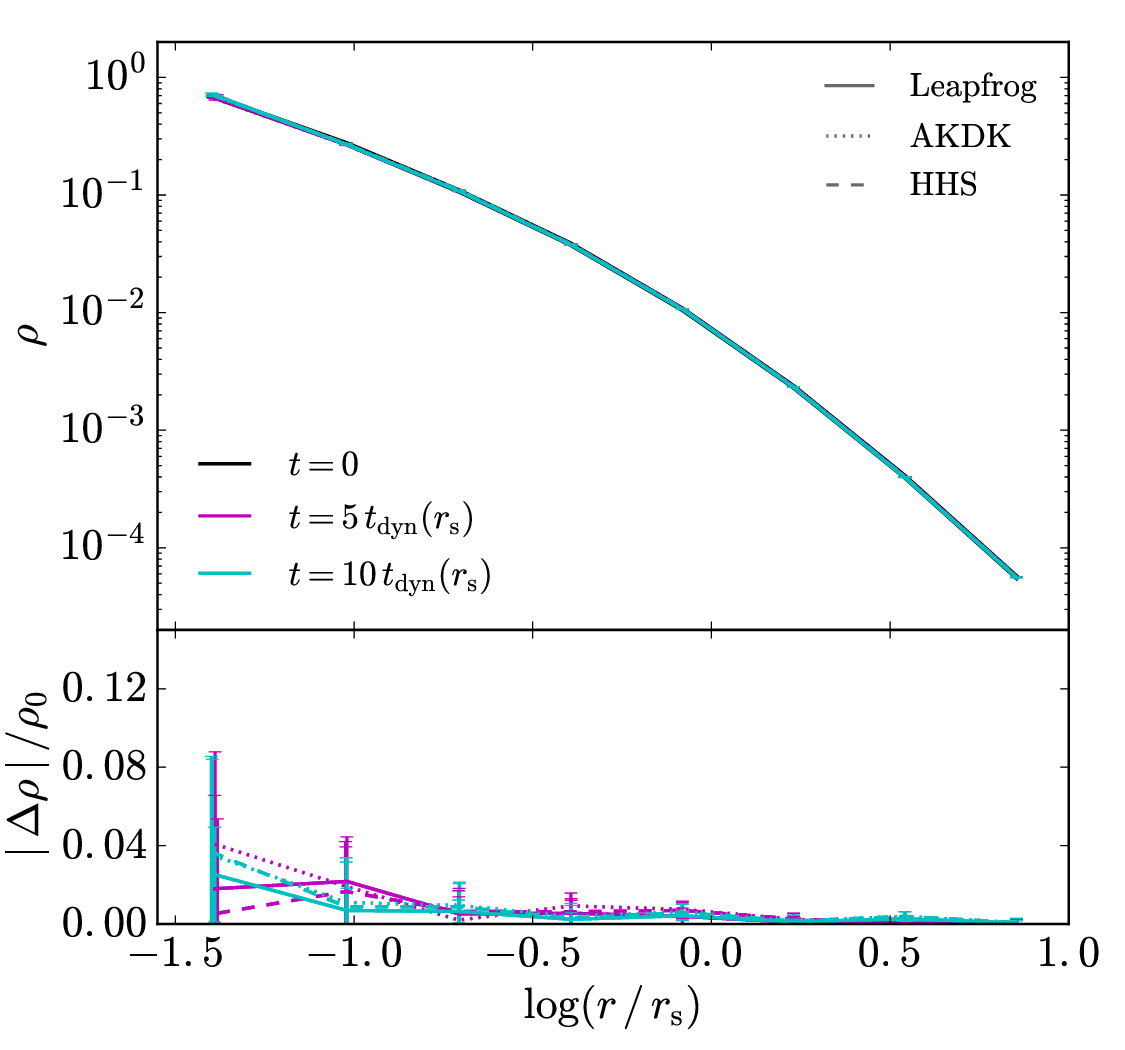}
    \caption{Density profile, at different times, of the isolated NFW halo with $2\times 10^6$ particles simulated up to 10 dynamical times for the three integrators: Leapfrog (solid lines), AKDK (dotted lines) and HHS (dashed lines). The three integrators accurately preserve the density profile.}
    \label{fig:HA_dens}
\end{figure}

%------------------------Figure----------------------------------%
\begin{figure*}
     \centering
     \begin{tabular}{cc}
        \includegraphics[width=0.9\textwidth,angle=0]{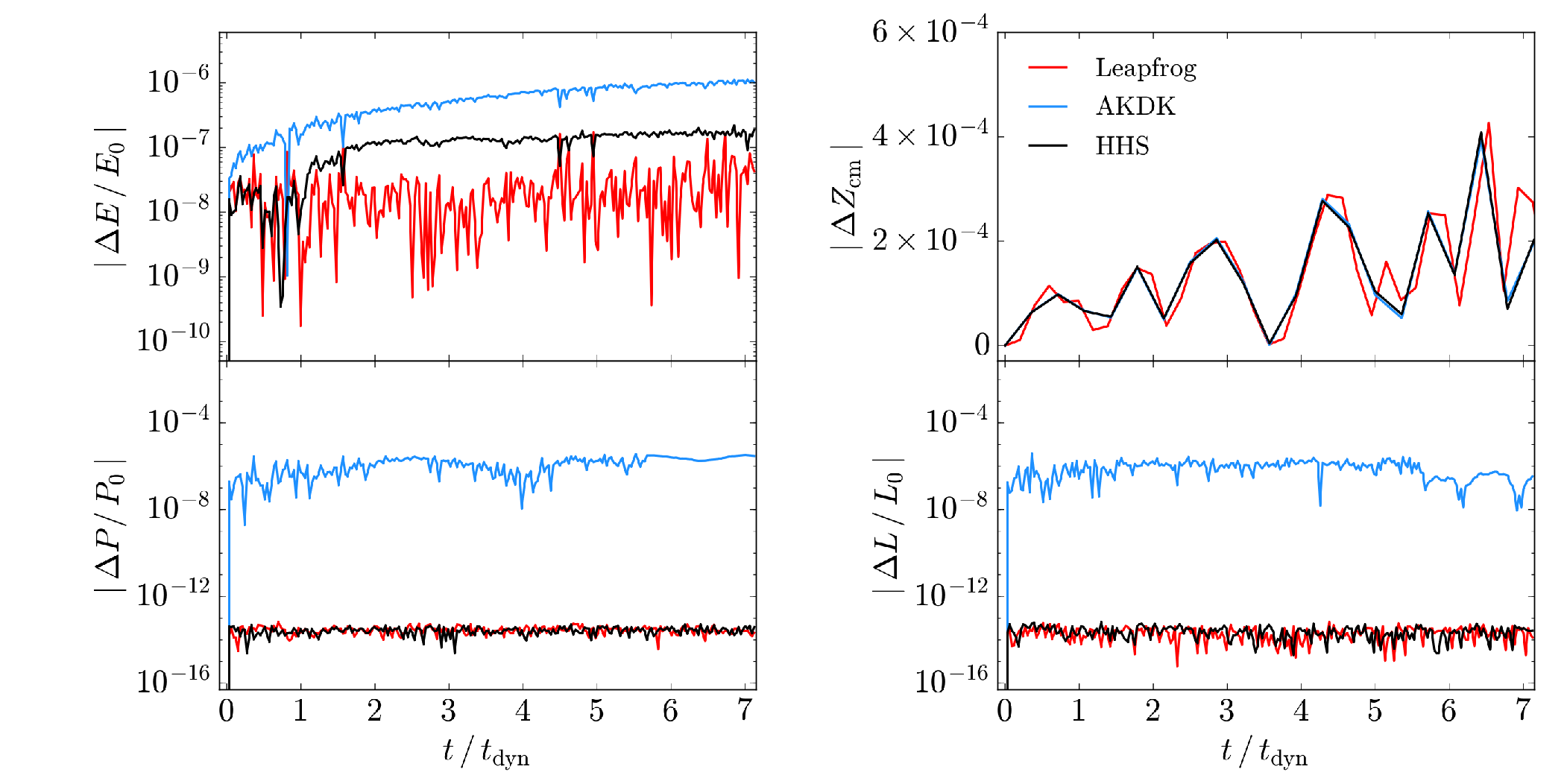}
     \end{tabular}
\caption{Error in conserved quantities for a NFW halo with $N=10^5$ particles plus a satellite simulated up to 7 dynamical times. Shown is the energy error (upper left panel), change in center of mass position (upper right panel), error in linear (lower left panel) and angular momentum (lower right panel) for the three different integrators: Leapfrog (red), AKDK (blue) and HHS (black).}
\label{fig:DF_ener}
\end{figure*}

Regarding the conservation of other dynamical quantities like linear and angular momentum or the system barycenter the situation is different (see figure \ref{fig:HA_ener_big}). Leapfrog and HHS preserve almost at machine precision the linear and angular momentum (see bottom panels), whereas AKDK presents smaller accuracy, although it has a slope until $10\,t_\mathrm{dyn}$, afterwards it flattens. Interestingly, HHS with a delay in updating the time-step hierarchy (HHS-sTSS, yellow) is almost indistinguishable from HHS and leapfrog. The upper right panel shows the accuracy in preserving the halo centroid. Once again, leapfrog and both HHS versions accurately keep the centroid, while AKDK accuracy degraded two orders of magnitude. The wall-clock time for leapfrog, HHS and HSS-sTSS experiments was, respectively, $1.7$, $0.8$ and $0.6$ times the corresponding for AKDK. As a complement, we performed tests using larger time steps reaching lower energy accuracy, the general results are the same. Finally, we emphasize that all integrators accurately preserve the density profile, as we can see in figure \ref{fig:HA_dens}; with obvious dependence on the number of particles, we decided to show the test with two million particles in order to minimize discreteness effects.

\subsection{Minor Merger: Sinking Satellites}

%------------------------Figure----------------------------------%
\begin{figure}
    \centering
    \includegraphics[width=0.47\textwidth,angle=0]{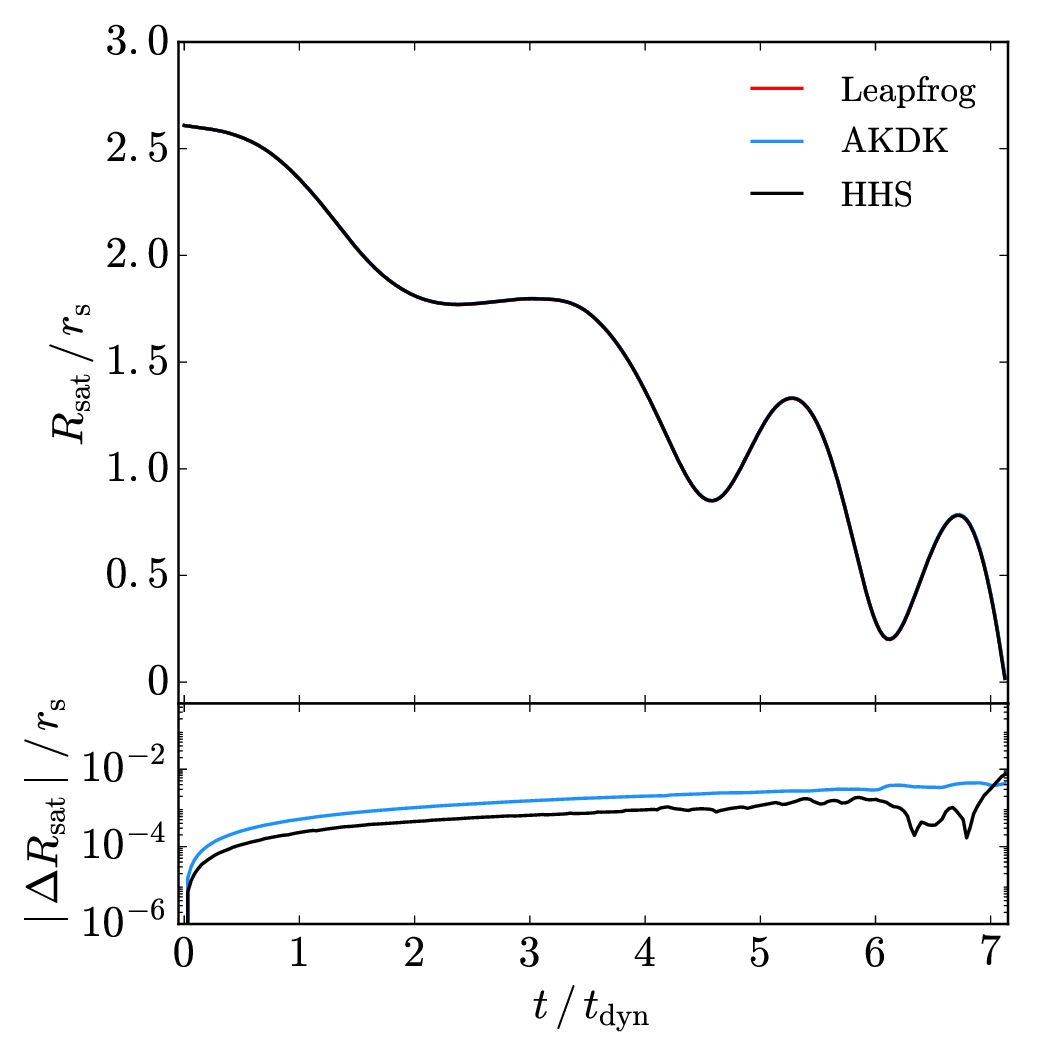}
    \caption{Evolution of the satellite radial position (upper panel) for the sinking satellite test and for the three integrators: leapfrog (red), AKDK (blue) and HHS (black). When comparing with leapfrog (lower panel), the differences are below $1\%$ between integrators.}
    \label{fig:Rsat_LF}
\end{figure}

Satellite accretion onto larger galaxies is an astrophysical problem commonly simulated by both, isolated and cosmological \emph{N}-body simulations \citep{2020MNRAS.495.4496M, 2016MNRAS.461.4335A}. We simulated a satellite, represented by a softened and massive particle, falling into a spherical system, represented by collisionless softened particles, for seven dynamical times. The spherical system consists of $N=10^{5}$ equal mass particles spatially distributed according to the NFW density profile \citep{1997ApJ...490..493N}. We adopted the same units as the previous fiducial isolated halo case. We used Plummer softening and we chose a softening parameter $\epsilon=0.026$ (in model units). The satellite's initial separation from the center of the spherical system is $R_{\mathrm{sat}}=2.6$, and its mass is $m_{\mathrm{sat}}=0.01$, which is $\sim$1300 times bigger than the mass of one collisionless particle.

This test is particularly useful because the sinking process involves orbital angular momentum and energy transfer into the host system. We tracked energy, linear and angular momentum conservation as well as the host center of mass behaviour (see figure \ref{fig:DF_ener}). Leapfrog (red) stays flat. HHS (black) starts to jump at one dynamical time, afterwards it stays flat. AKDK (blue) shows a lower accuracy in energy conservation and presents a systematic energy growth. As in the case of the isolated halo, HHS energy drift slope is considerably flatter than the corresponding to AKDK. Linear and angular momentum are less accurate for AKDK for several orders of magnitude, while system barycenter behaves essentially the same for all three integrators. These results indicate that HHS is an excellent alternative for dynamical friction studies.

The evolution of the satellite radial position shows differences below the one percent level (see figure \ref{fig:Rsat_LF}). We conclude that, for a reduced number of dynamical times, the three integrators can provide an accurate description of the sinking process.

\section{Time-step selection function tests} \label{TS_tests}

As it has already been discussed some hierarchical/adaptive time-step integrators, like HHS and AKDK, include a time-step selection function, such a prescription may help restoring the integrator symmetry. We may question if the approximated symmetric time-step selection function based on particle pairs given by eq. \ref{eq:symmHS} is only useful for a particular kind of simulation or code, and if we loose all the convenient properties of HHS, observed at this point, when the problem is not tractable by a direct summation code. To quantify such effect we evolved the fiducial isolated halo switching the time-step selection function and we additionally performed test with the Tree/FMM code GADGET4.

\subsection{Direct summation code} \label{TSsF_PP}
\begin{figure*}
    \centering
    \includegraphics[width=0.53\textwidth,angle=0]{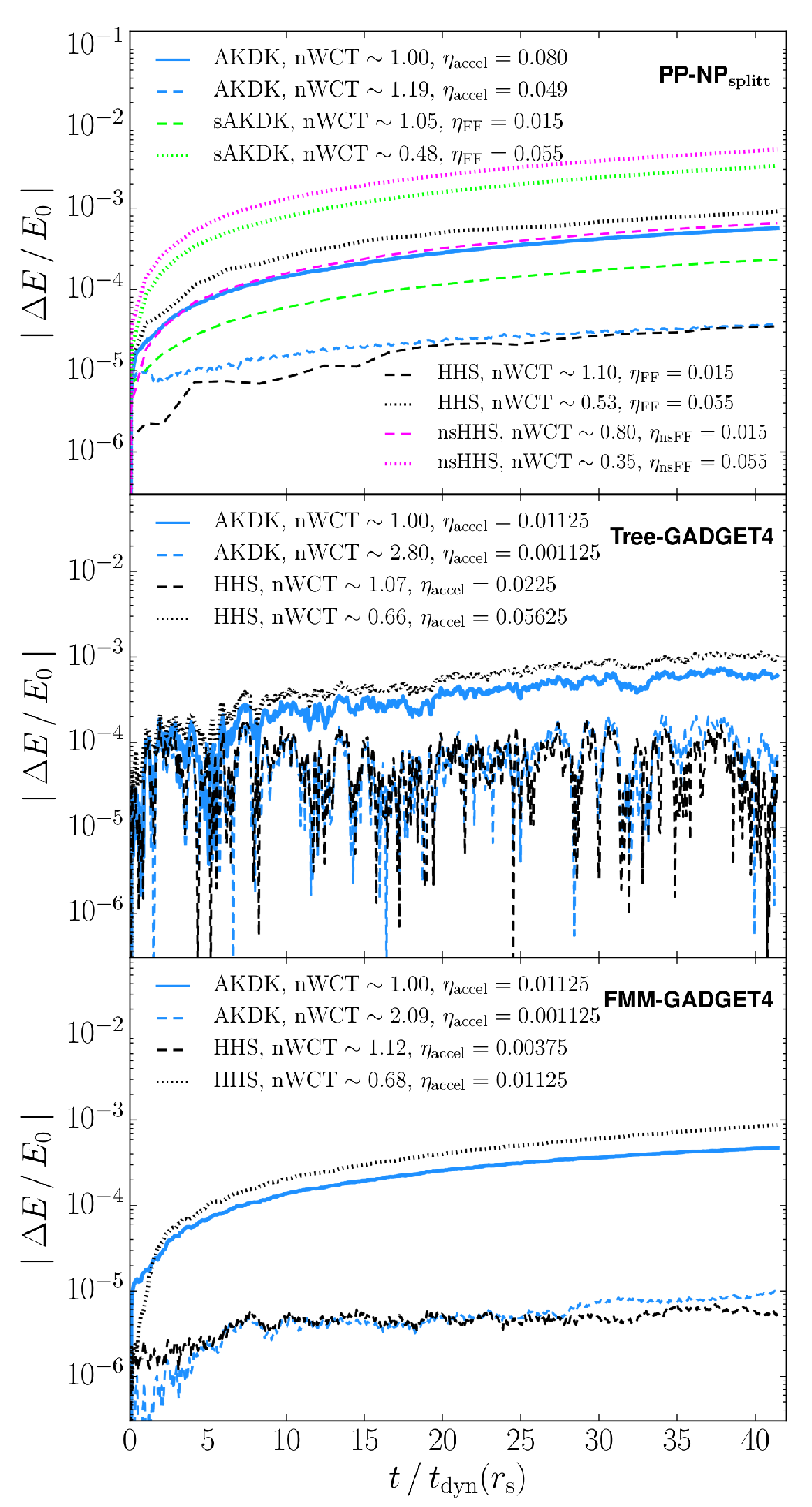}
    \caption{Effect of time-step selection function and different acceleration codes. nWCT corresponds to the wall-clock time of each test normalized to the fiducial AKDK (solid blue line). Upper panel shows tests with our direct summation code. Solid blue line is the fiducial AKDK test and dotted black line is HHS, almost 50$\%$ faster. Dashed lines correspond to experiments with smaller accuracy parameter but reaching a limit where HHS is faster and more accurate than AKDK. Magenta dotted/dashed lines show the same HHS tests but neglecting the symmetrizing derivative term in eq. \ref{eq:symmHS}, even going with smaller $\eta_\mathrm{nsFF}$ still may be competitive with AKDK. Green dotted/dashed lines show AKDK test with the symmetrized time-step selection function (eqs. \ref{eq:symmHS} and \ref{eq:freefall}) using the same parameters as HHS cases (black dotted/dashed lines, respectively). Middle panel shows experiments with GADGET4 using the Tree version. Results are consistent with the direct summation code. Solid blue line is the fiducial AKDK, dotted black line is HHS $7\%$ slower and slightly less accurate. Dashed lines are AKDK and HHS, with almost flat behaviour, slightly better for HHS, however, AKDK is almost three times slower. Lower panel shows the equivalent tests but now for FMM GADGET4. As before, the AKDK for the flat case (dashed blue) is slightly worse in energy accuracy and almost twice as slow as compared with HHS (dashed black). We conclude that, even for different time-step functions, there is still a regime where HHS is more efficient.}
    \label{fig:ener_sFunc}
\end{figure*}

First we consider our direct summation code implementation, simulating the fiducial isolated halo considering: HHS and AKDK integrators, as it can be seen in the upper panel of figure \ref{fig:ener_sFunc}. We take as a reference the AKDK integrator with an accuracy parameter $\eta_\mathrm{accel} = 0.08$ (fiducial, solid blue line). The dotted black line shows HHS (with $\eta_\mathrm{FF}=0.055$), which is almost twice faster but slightly less accurate, and the dashed black line shows a case of HHS (with $\eta_\mathrm{FF}=0.015$) remarkably more accurate but $10\%$ slower than the fiducial AKDK. We include another AKDK case (dashed blue) with a smaller accuracy parameter $\eta_\mathrm{accel} = 0.049$, which is $20\%$ less accurate than HHS (dashed black line) but $10\%$ slower, and also is $20\%$ slower and an order of magnitude more accurate than the fiducial AKDK case (solid blue). This means that considering even smaller values for the parameter $\eta$ will not make AKDK more efficient that HHS. Next, we perform some changes in the time-step selection function, we removed the derivative term from eq. \ref{eq:symmHS} which represents the symmetrizing correction for the HHS integrator, we dubbed such tests nsHHS and they appear as the magenta lines in figure \ref{fig:ener_sFunc}, the energy drift is larger as compared with HHS but it is still acceptable for collisionless simulations and it is faster. For completeness, we performed tests using eq. \ref{eq:symmHS} in AKDK (dubbed as sAKDK, green lines) with the same parameters ($\eta_\mathrm{FF}$, rungs, etc.) as in HHS tests. The energy drift is larger for the sAKDK test than the one corresponding to HHS, however sAKDK is faster. Note that it is possible to match HHS accuracy by lowering the $\eta_\mathrm{FF}$ parameter in sAKDK tests, but it is slower (we do not show it because it does not add new information). So far, our tests suggest that HHS may benefit hybrid codes that use direct summation force calculation as part of their algorithm, for example in the P3M technique \citep{1988csup.book.....H} the most expensive part corresponds to direct summation, which runs in accelerators like GPUs \citep{2013hpcn.confE...6H,2016NewA...42...49H, 2020arXiv200303931C}. Even using different time-step selection function, HHS may still obtain considerable performance. We analyze such a situation in the following sections, as well as, HHS performance in codes using approximated force computations (e.g. a tree code).

%------------------------Figure----------------------------------%

\subsection{Tree Code: GADGET4}
Recently, the 4th version of the publicly available code GADGET has implemented the HHS in the so called hierarchical gravity mode \citep{2021MNRAS.506.2871S}. The time-step selection function is similar to our equation \ref{eq:standard} but the accuracy parameter $\eta_\mathrm{accel}$ is inside the square root. Changing the step function is a sensible choice because our equation \ref{eq:symmHS}, based on pair interactions, is not practical for very large number of particles. Also, in the GADGET4 HHS implementation, instead of using the DKD representation (as in our implementation), the authors adopted the KDK one \citep[for further details we refer to][]{2021MNRAS.506.2871S}, in a similar way as \citet{2017arXiv171210116Z}. Hence, this allow us to explore the case of an approximated gravitational acceleration code, such as tree code \citep[e.g.][]{1986Natur.324..446B}, with different time-step selection function. For that purpose, we perform some tests using AKDK and HHS with the Tree version of GADGET4. As in the case of direct summation tests (section \ref{TSsF_PP}), we simulated the fiducial isolated halo in such a way we can directly compared with the direct summation code tests, including a non-symmetrized time-step function. Middle panel of figure \ref{fig:ener_sFunc} shows the results for the Tree code version of GADGET4. For the corresponding fiducial AKDK (solid blue line), GADGET4 opens three time-step rungs, and for its HHS implementation (dotted black line), it opens only two time-step rungs. Both integrators show an energy drift similar to $10^{-3}$, however, HHS (dotted black line) is $66\%$ faster but less accurate. Motivated for such result, we run a new HHS test decreasing the $\eta_\mathrm{accel}$ parameter (dashed black line), the energy is considerably flat, therefore energy conservation after $40\,t_\mathrm{dyn}$ is almost an order of magnitude better than the fiducial AKDK (solid blue line), but it is $7\%$ slower. If we decrease $\eta_\mathrm{accel}$ for the AKDK integrator in order to match energy conservation (dashed blue line), the wall-clock time is considerably larger than $50\%$. Although this is a particular example model, it is consistent with our previous tests. HHS is more stable and at medium and long-term integrations may be more efficient than AKDK, regardless of not using the approximated symmetric free-fall particle pairs time-step function (eqs. \ref{eq:symmHS} and \ref{eq:freefall}).

\subsection{Fast Multipole Method Code: GADGET4}
Gadget4 has implemented a different gravity solver based on the Fast Multipole Method expansion \citep[FMM, e.g.][]{2000ApJ...536L..39D,2002JCoPh.179...27D}. We run the same fiducial isolated halo, as in previous section, adopting the FMM scheme truncating the expansion until the quadrupole term. Results are presented in the lower panel of figure \ref{fig:ener_sFunc}. Blue solid line shows the fiducial case using AKDK ($\eta_\mathrm{accel}=0.01125$), the corresponding HHS case ($\eta_\mathrm{accel}=0.01125$, dotted black line) is relatively faster ($66\%$), however, it is less accurate. We experimented lowering the $\eta_\mathrm{accel}$ parameter for HHS, the energy evolution is flat (dashed black line). We also decrease $\eta_\mathrm{accel}$ for AKDK (dashed blue line), the energy accuracy is $10\%$ worse but the wall-clock time is almost $100\%$ larger; therefore, there is no point trying smaller $\eta_\mathrm{accel}$ values. The experiments with GADGET4 FMM version also confirm the conclusions from our direct summation tests.

\section{Performance} \label{Perfomance}

%------------------------Figure----------------------------------%
\begin{figure}
    \centering
    \includegraphics[width=0.46\textwidth,angle=0]{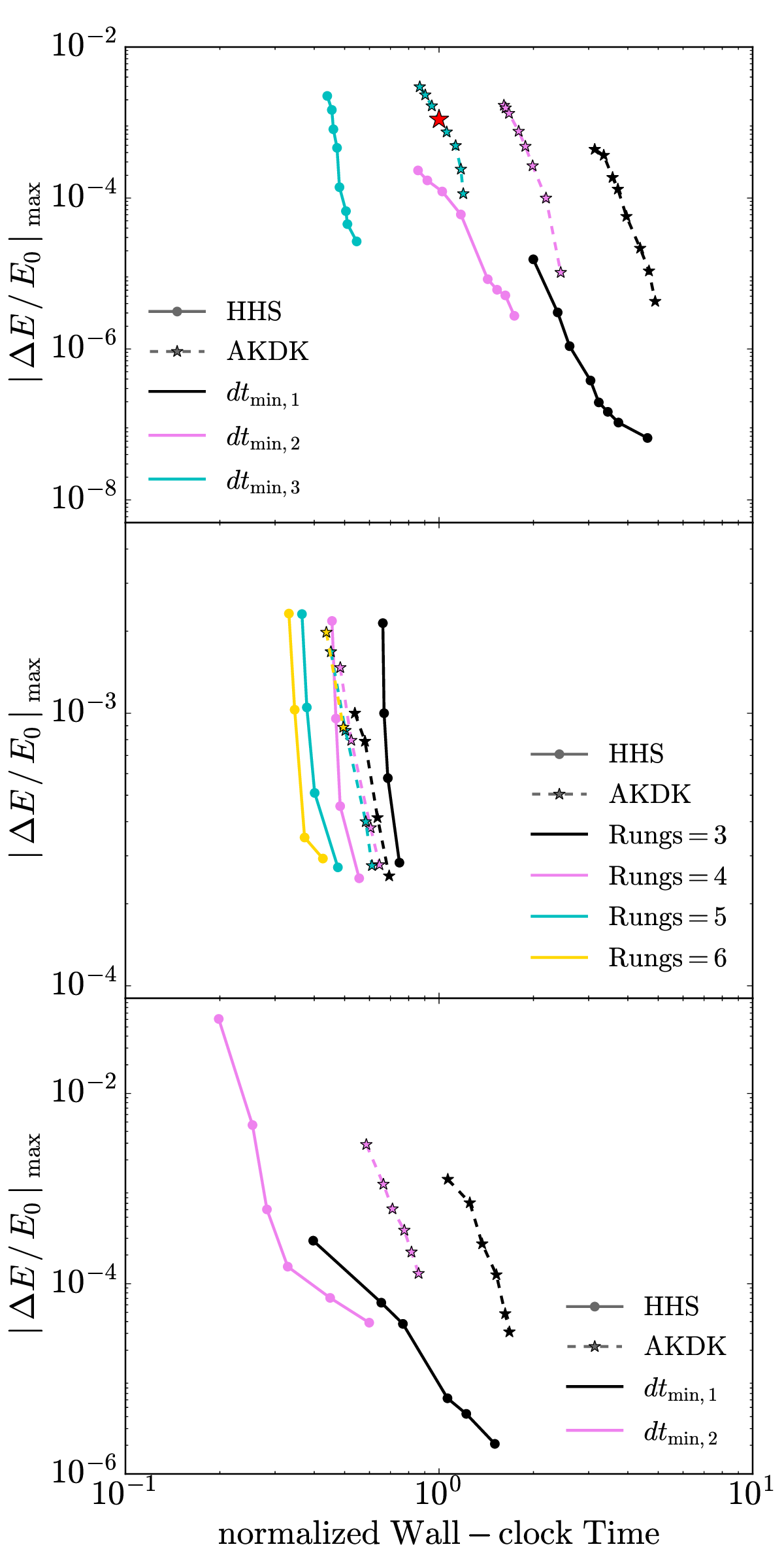}
    \caption{Energy conservation as a function of wall-clock time (normalized with respect to the red star case in upper panel) for the fiducial isolated halo (upper and middle panels) and for a cuspy halo including four live cuspy satellites (bottom panel), for HHS (solid circles) and AKDK (stars) integrators. In upper panel, we used a plummer softening $\epsilon = 0.007$, and we varied the minimum time step ($dt_\mathrm{min,1}=5.4\times10^{-3}$, $dt_\mathrm{min,2}=1.1\times10^{-2}$ and $dt_\mathrm{min,3}=2.2\times10^{-2}$) but fixing the number of rungs ($6$ and $5$ for HHS and AKDK, respectively). In middle panel, we used $\epsilon = 0.01$, and we varied the number of rungs but fixing the minimum time step ($dt_\mathrm{min}=4.3\times10^{-2}$). For the bottom panel, we used $\epsilon = 0.004$, and we varied the minimum time step ($dt_\mathrm{min,1}=6.4\times10^{-3}$ and $dt_\mathrm{min,2}=1.3\times10^{-2}$) but fixing the number of rungs ($6$ and $3$ for HHS and AKDK, respectively).}
    \label{fig:pres_wct}
\end{figure}

%------------------------Figure----------------------------------%
\begin{figure}
    \centering
    \includegraphics[width=0.47\textwidth, angle=0]{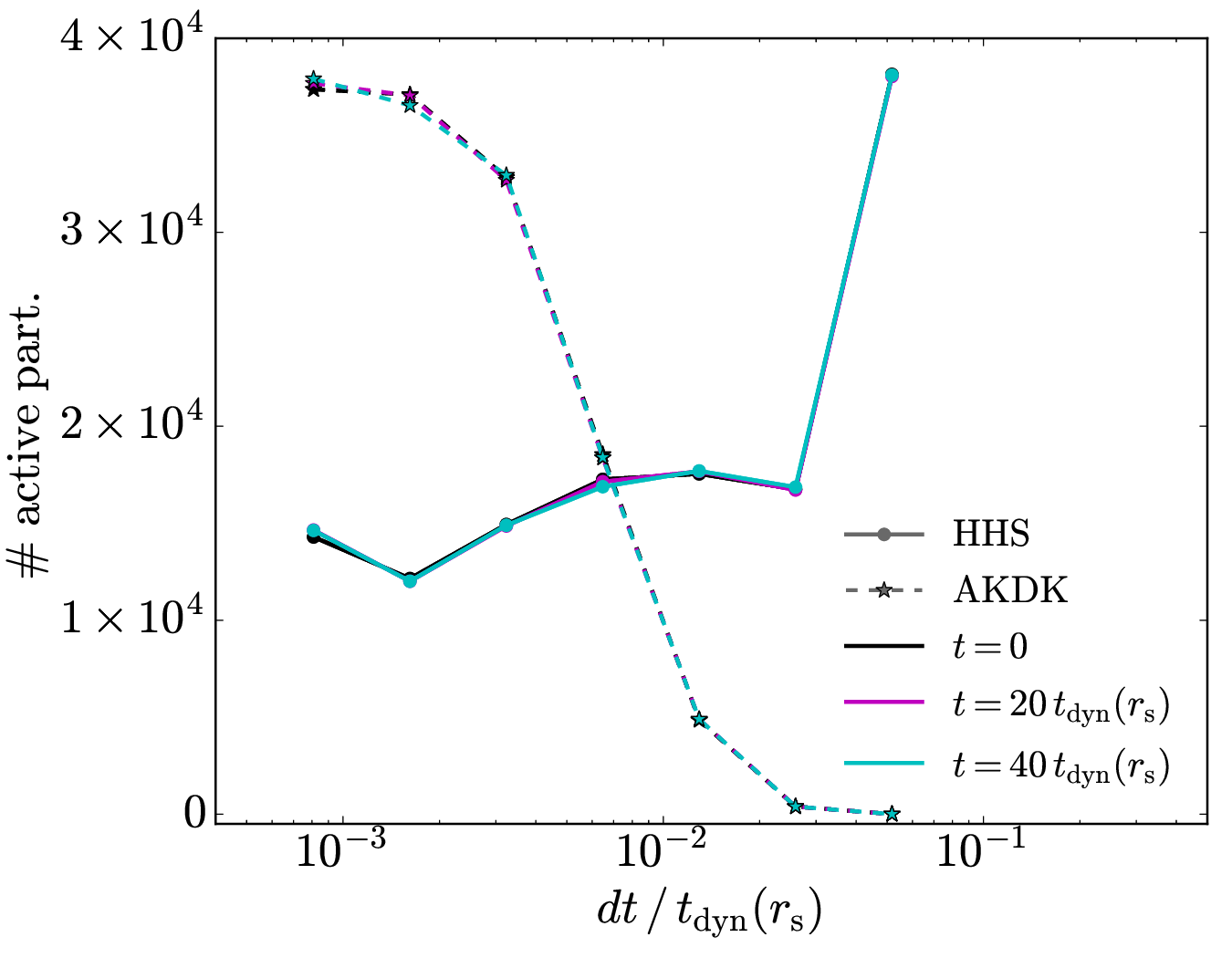}
    \caption{Particle distribution across time-step rungs for different selection functions: acceleration criteria (eq. \ref{eq:standard}, stars) and the approximated symmetric free-fall (eqs.\ref{eq:symmHS} and \ref{eq:freefall}, solid circles), for the fiducial isolated halo using the same parameters as figure \ref{fig:HA_ener_big}. It is notorious that the acceleration criteria has more particles with small time step and that free-fall criteria is almost the opposite, with potential consequences for the performance.}
    \label{fig:histo_dtis}
\end{figure}

Although the adaptive nature of the HHS and AKDK integrators may imply a higher efficiency compared with the global-constant time-step leapfrog, the benefit of taking adaptive time-steps is evident only when the dynamical range is large. In comparison with AKDK, HHS has a computational overhead due to the recursive splitting of the Hamiltonian needed to build the time-step hierarchy.

We decided to make a short exploration of the simulation parameters (accuracy, rung number, and minimum time-step) with our direct summation code. We present the results in figure \ref{fig:pres_wct}, which shows a pragmatic diagnostic of the integrator performance: energy conservation vs. wall-clock time. The upper and middle panels correspond to the fiducial isolated halo. In the upper panel we fixed the number of time-step rungs and we varied the minimum time-step; whereas in the middle panel, we fixed the minimum time-step (same for both integrators) and we varied the number of time-step rungs. The lower panel corresponds to the cuspy halo including four live cuspy satellites, a common situation in cosmological simulations that requires a large dynamical range. In many cases HHS outperforms AKDK. We may wonder what is the reason given the extra computations related with the splitting process. One possible suspect is the individual time-step distribution. To investigate that, we built the histogram of time steps for particles in the initial conditions and at later times (see figure \ref{fig:histo_dtis}). For HHS (solid circles) it is clear that only a moderate fraction of particles are found in the deepest time-step rung. For AKDK the distribution is almost the opposite, there is a peak of particles in the three deepest rungs, which translates into many more time-steps than HHS. The difference in performance is partly due to the time-step selection function, in agreement with \citet{2007MNRAS.376..273Z, 2020MNRAS.495.4306G}.  Although tweaking parameters we may obtain differences in performance and accuracy, we noticed that for a defined energy conservation threshold in many cases HHS outperforms AKDK, because the energy growth is smaller for HHS than for the standard AKDK implementation, allowing HHS to use larger steps. Nevertheless it is important to mention that if all the particles are similarly distributed in the time-step rungs for both integrators the computational overhead of HHS starts to play a more important role, this may happen for very low resolution runs where the dynamical range is artificially shortened.  This is in agreement with the middle panel of figure , showing that only when using more than three step runs HHS is faster than AKDK. The time-step histogram presented in figure \ref{fig:histo_dtis} is a handy tool to asses the situation.

Our conclusions are consistent with recent studies that published successful results using HHS with GADGET4 and reaching an extremely large dynamical range with multi-million particle numbers \citep{2020Natur.585...39W}.

\section{Long-term Stability} \label{Stability}

%------------------------Figure----------------------------------%
\begin{figure*}
    \centering
    \includegraphics[width=0.9\textwidth,angle=0]{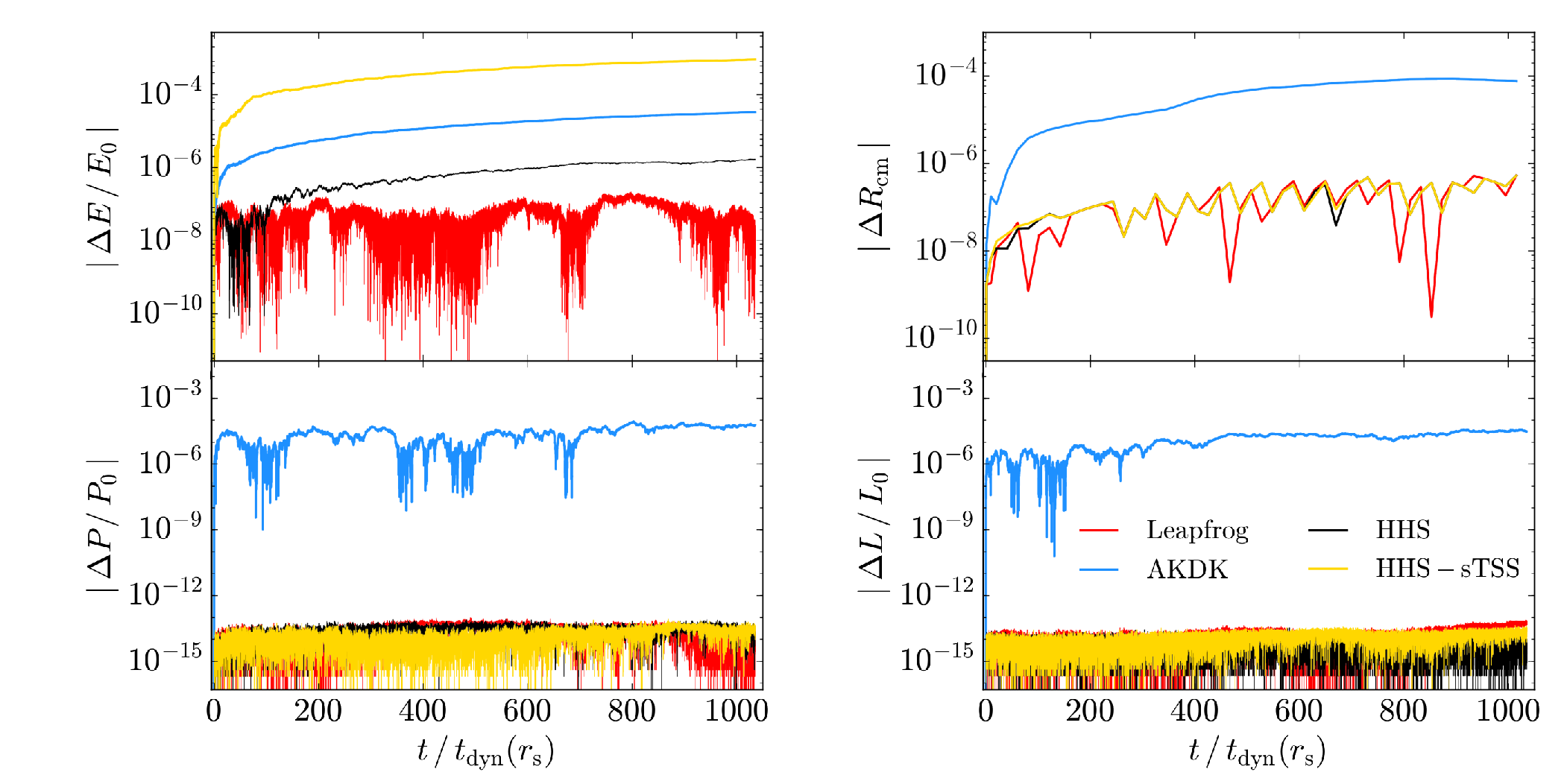}
    \caption{As figure \ref{fig:HA_ener_big}, but following the fiducial isolated halo up to $10^{3}$ dynamical times (at scale radius). }
    \label{fig:HA_long}
\end{figure*}

%------------------------Figure----------------------------------%
\begin{figure}
    \centering
    \includegraphics[width=0.47\textwidth,angle=0]{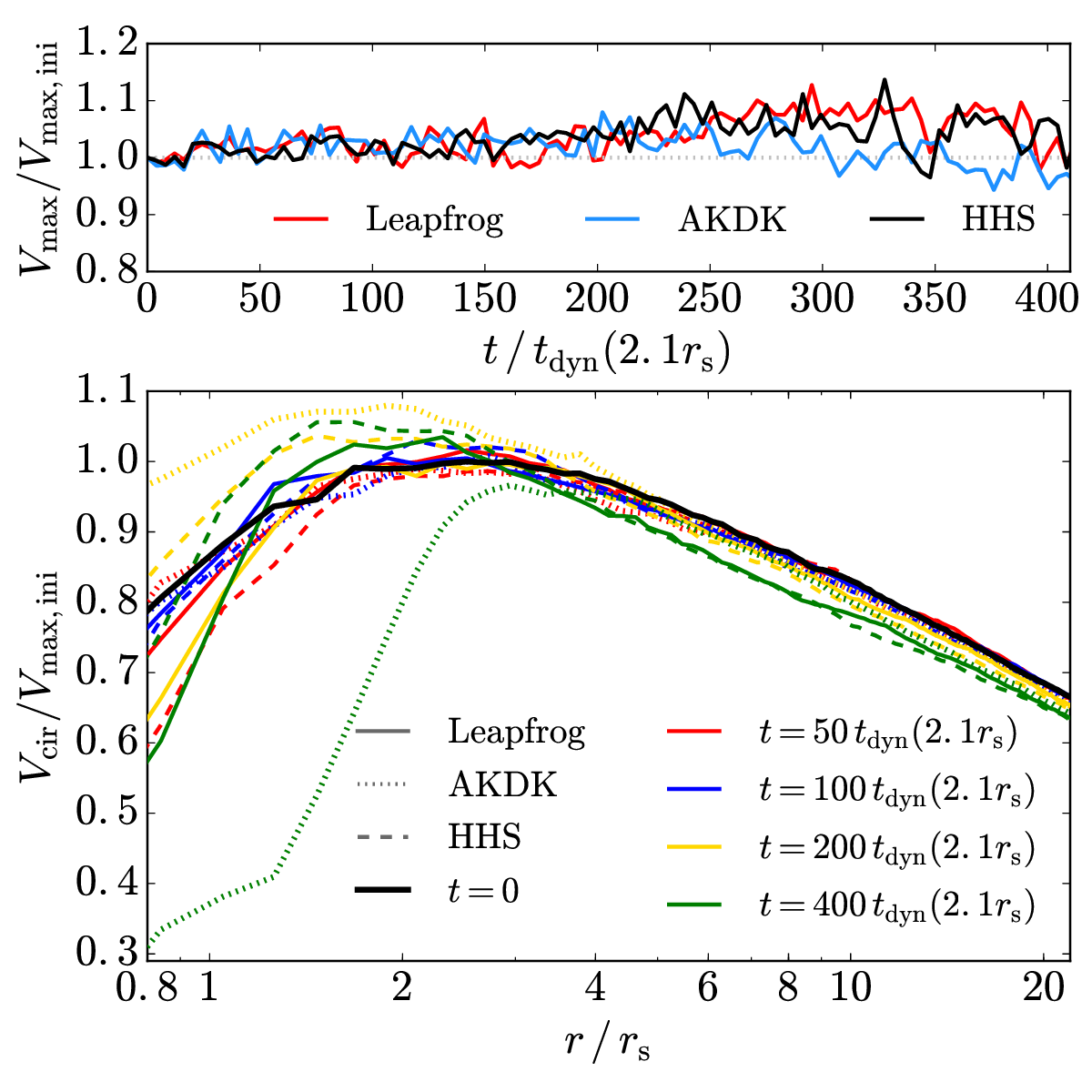} 
    \caption{Circular velocity curves for the fiducial isolated halo with $2000$ particles following the system for long integration times. There is not systematic difference between different integrators. The energy conservation level is $\sim 10^{-5}$. The maximum circular velocity is scattered inside $10\%$ at different moments.}
    \label{fig:Vc}
\end{figure}

%------------------------Figure----------------------------------%
\begin{figure*}
    \centering
    \includegraphics[width=0.77\textwidth,angle=0]{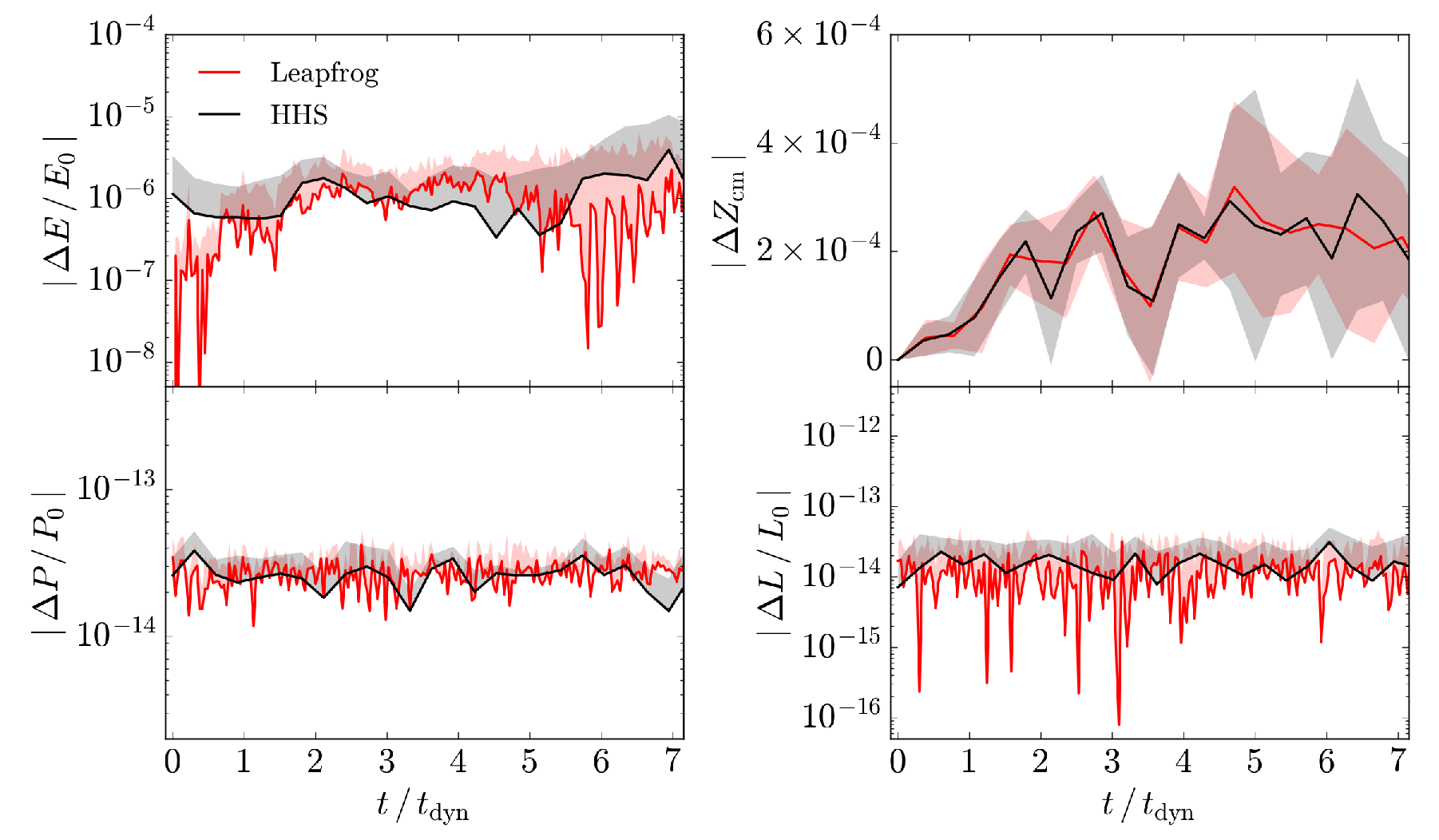}
    \caption{Error in conserved quantities for a NFW halo with $N=10^5$ particles plus a satellite simulated up to 7 dynamical times. Shown is the energy error (upper left panel), change in center of mass position (upper right panel), error in linear (lower left panel) and angular momentum (lower right panel) for the Leapfrog (red) and HHS (black) integrators. Shaded portions represent one sigma standard deviation propagated from different realizations run under different random seeds. }
    \label{fig:reali_DF_ener}
\end{figure*}

At this point we have compared the accuracy and performance of the integrators for some dynamical times. A natural question arises about if HHS advantages are relevant for realistic long-term integration. Dark matter halos survive around $30-200$ dynamical times in cosmological simulations depending on the merger/accretion history \citep{2015MNRAS.447.3693K}. As before we investigated the stability of energy, linear and angular momentum and density centroid for our fiducial isolated halo model, this time for hundreds of dynamical times (see figure \ref{fig:HA_long}). Energy conservation of HHS (black) is quite close to the global-constant time-step leapfrog (red) behaviour during the first $20-40$ dynamical times (consistent with our previous tests), however, after that it starts to slowly drift which seems to get smaller at the end of the simulation ($\sim 700\, t_\mathrm{dyn}$). AKDK (blue) quickly drifts to a considerably larger energy error and keeps systematically growing. The yellow curve represents the HHS-sTSS version that delays the time-step hierarchy updating, as we observed before, it behaves as AKDK but with smaller accuracy, although it is faster ($\sim 40\%$). For linear and angular momentum all integrators preserve such quantities, however, while HHS preserves them almost at machine precision, AKDK preservation is almost $8$ orders of magnitude worse. A similar disparity is obtained by following the halo centroid. Seeking for a consequence of this difference in accuracy we tracked the circular velocity in experiments with a smaller number of particles ($N=2000$) evolving the system for $400$ dynamical times at $2.1 r_\mathrm{s}$, where circular velocity peaks. Figure \ref{fig:Vc} shows the circular velocity profile at several times. Despite we find some differences, they are all inside $10\%$. We conclude that, at the level of high accuracy, we do not expect important differences between integrators and performance is the most relevant difference. As far as \emph{N}-body simulations reach a larger dynamical range, the smallest structures may live for a larger number of dynamical times, in this context HHS may offer a more stable option.

Recent studies regarding long-term \emph{N}-body evolution \citep{2020MNRAS.493.1913H}, suggest that long-term integrations may be unstable to small changes in initial conditions realizations, this may be particularly critical for highly non-linear situations like the three-body problem. We generated a small ensemble of realizations for the fiducial isolated halo and for the sinking satellite problem, results are shown in figure \ref{fig:reali_DF_ener} only for Leapfrog (red) and HHS (black). Indeed some scatter is found, however, overall the results are robust.

\section{Reversibility and time symmetry} \label{Reversibility}

%------------------------Figure----------------------------------%
\begin{figure*}
    \centering
    \includegraphics[width=0.65\textwidth,angle=0]{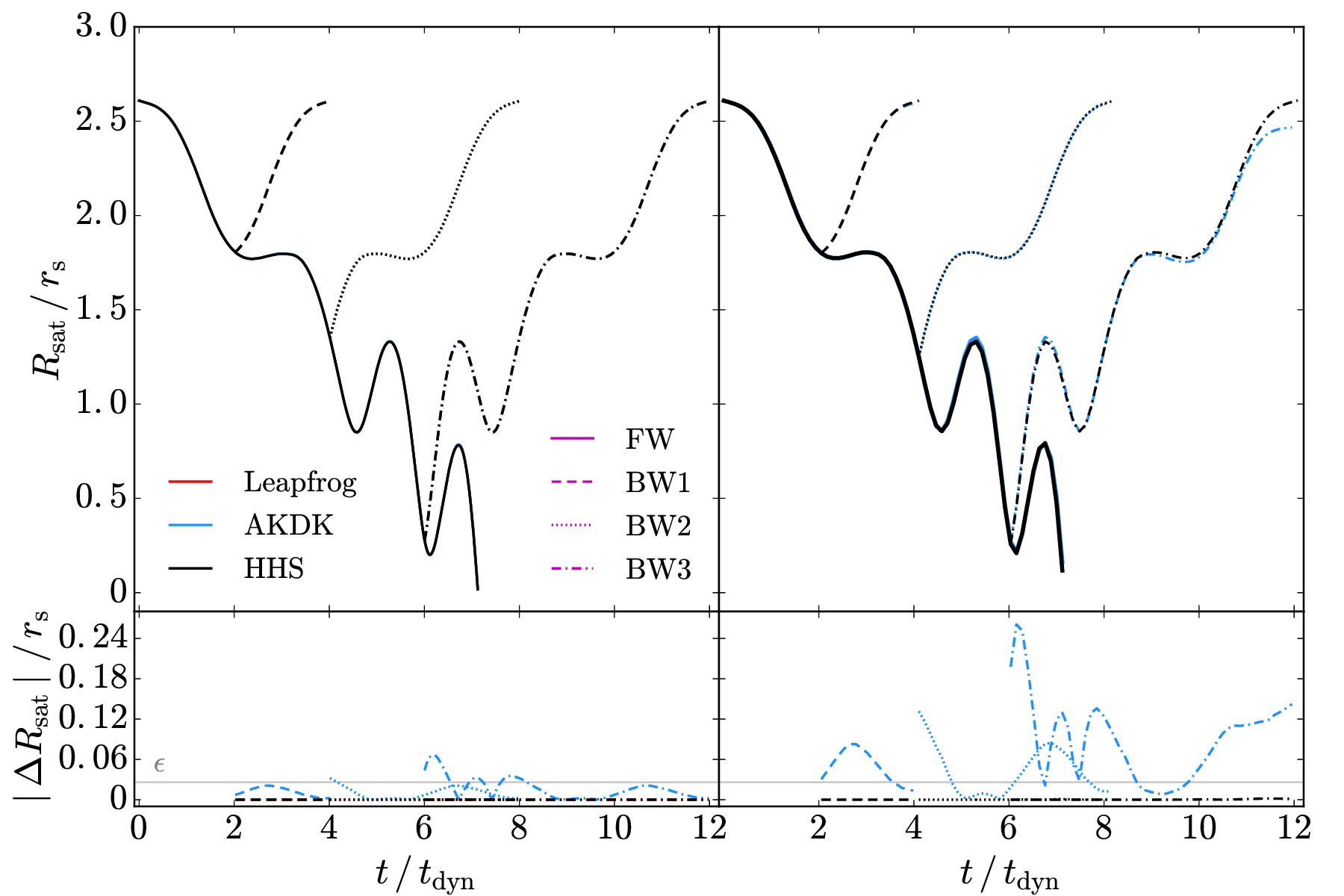}
    \caption{Reversibility test. For the sinking satellite experiment we reversed the velocity sign at different moments, indicated by BW1 (dashed), BW2(dotted) and BW3 (dash-dot), and continue the integration. Solid lines represents the forward integration. Purple line labels only indicate the line-style corresponding to the start of each FW/BW integration. Upper panels show the normalized satellite distance as a function of time for the truly symplectic leapfrog (red lines), the adaptive one AKDK (blue lines) and HHS (black lines) integrators. Lower panels show the difference in position regarding the corresponding forward solution (solid lines). For high energy conservation accuracy ($\sim 10^{-8}$, left panels), there are some differences between integrators, however, they are inside (Leapfrog and HHS) or very close (AKDK) to the simulation softening (horizontal grey line). For a lower energy conservation accuracy ($\sim 10^{-3}$, right panels), still HHS and Leapfrog position differences are inside the simulation softening, whereas AKDK case is above the simulation resolution. }
    \label{fig:rev_sat}
\end{figure*}

The above numerical experiments show that there are certain parameter combinations where HHS is more accurate than AKDK or, alternatively, it is faster for a given energy accuracy. Because both integrators have different parameters it is natural to ask if there is a true advantage of HHS or if it is a misleading result,  dependent on our implementation, accuracy or even particle number.

As we discussed in the introduction and in agreement with  \citet{2003AcNum..12..399H} we performed time symmetry and velocity reversibility tests using both HHS and AKDK integrators,  we used leapfrog with a global-constant time-step as a reference case.  

\begin{figure}
    \centering
    \includegraphics[width=0.47\textwidth,angle=0]{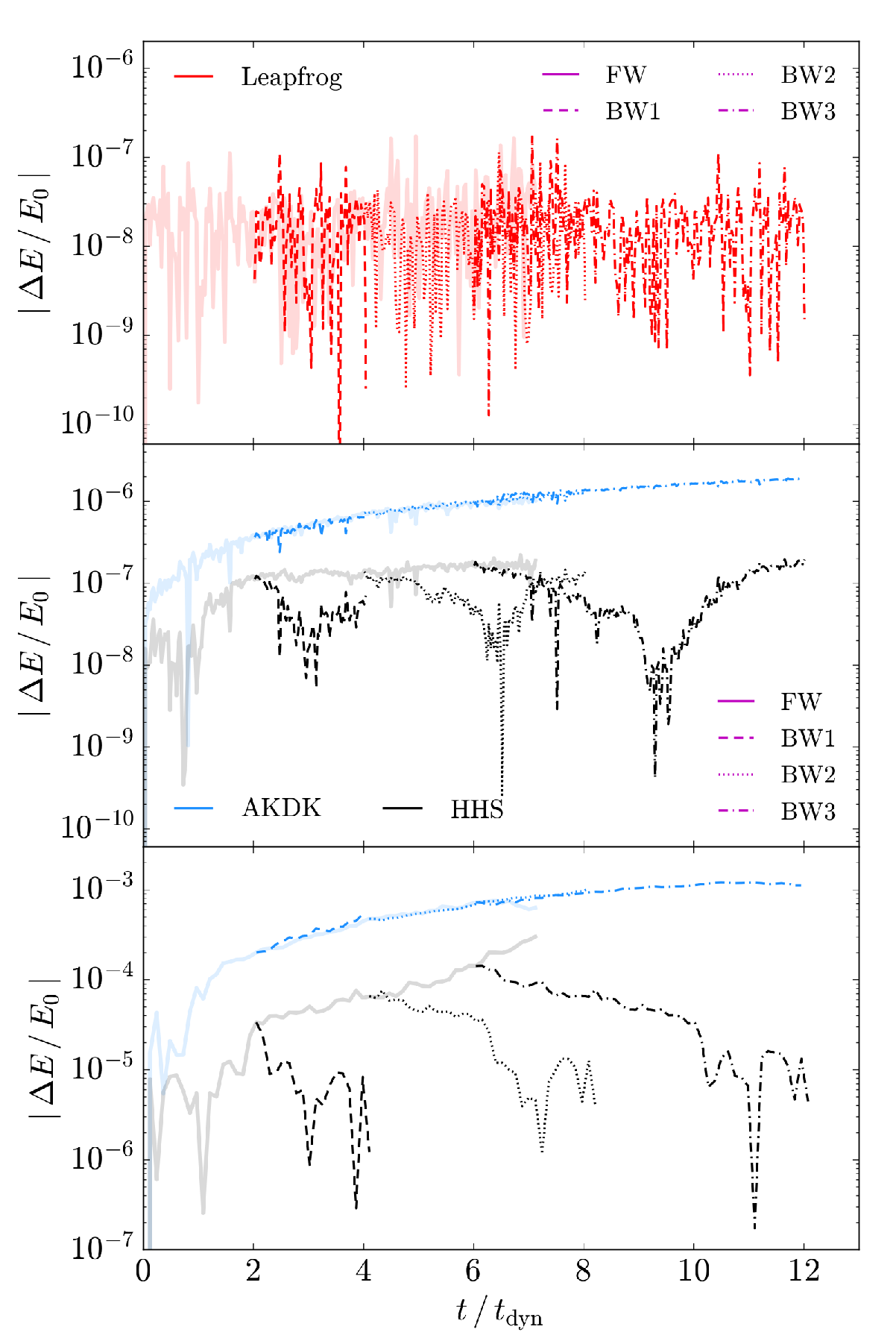}
    \caption{Fractional energy error before (solid opaque lines) and after reversibility (dashed, dotted and dash-dot lines for BW1, BW2 and BW3 moments, respectively), for the same test of figure \ref{fig:rev_sat}. Top panel shows the Global-constant time-step leapfrog, middle panel shows the higher accuracy in energy conservation tests, and the bottom panel the lower accuracy tests. Leapfrog is truly reversible. AKDK (blue lines) has a systematic growth of energy, suggesting that it is non-reversible. HHS (black lines) is almost flat with a small drift of energy after $\sim 2\,t_\mathrm{dyn}$ ($4.5\, t_\mathrm{dyn}$) for the high (low) accuracy test of the backward integrations, indicating that it is approximately reversible. Note that the purple line labels only indicate the line-style corresponding to the start of each FW/BW integration. }
    \label{fig:rev_ener}
\end{figure}

We chose the sinking satellite system as the test bed because the satellite orbit allows easily to track the system response in configuration space as well as in energy. At three different moments, termed BW1, BW2 and BW3, we reversed velocity signs for all particles, and for one case instead of velocities we reversed time sign, after that we continued the integration. Figure \ref{fig:rev_sat} shows the global result of forward (FW, solid lines) and backward (BW) evolution (i.e. inverting the sign of velocities) of the satellite distance to the halo center for all integrators either for high (left panels) or low (right panels) energy conservation accuracy. For high energy conservation accuracy ($\sim 10^{-8}$, left panels), the differences in configuration space are in general small, which is consistent with \citet{2018MNRAS.475.5570H}, where they discuss that AKDK preserves quantities like angular and linear momentum (see also figure \ref{fig:DF_ener}), however, there are still some differences as we can see in the lower panel. The satellite distance change with respect to its own forward evolution (lower panels) is a good diagnostic of reversibility. Clearly, the position difference in HHS (black) regarding the symplectic Leapfrog (red, global constant step) is below the simulation resolution determined by the softening (horizontal grey line), while for AKDK (blue) the calculated position  difference is larger but close to the simulation softening. At a less accurate  energy conservation but more common in collisionless simulations ($10^{-3}$, right panels), differences between forward and velocity reversal integration in Leapfrog (global constant step) and HHS are still below the simulation softening, however AKDK has notorious differences that are well above the simulation resolution and they are even detected in configuration space. We also tracked the fractional energy change (figure \ref{fig:rev_ener}). The first outstanding fact is the truly reversible behaviour of the global-constant time-step leapfrog (upper panel), almost every peak and valley is reproduced after the velocity reversal. The middle panel shows the high accuracy tests for AKDK (blue lines) and HHS (black lines), and the bottom panel the corresponding low accuracy tests. For AKDK, there is a roughly systematic growth for both forward integration (blue solid opaque line) and also after reversing velocities, suggesting that AKDK is non-reversible. Instead, HHS is almost flat, suggesting that it is approximately reversible. We also run a test using HHS where we change the time variable sign. In this case, the energy shows a systematic growth, showing that HHS is not time symmetric, however, for the sake of clarity, it was not included. We performed similar tests for the fiducial isolated halo and obtained similar results. The small slope showed in the energy drift suggests that HHS is approximately reversible, furthermore energy drift is small compared to AKDK that is no-reversible and non time-symmetric. This in agreement with \citet{2002Springer..} who shows that the symplectic St\"ormer-Verlet-LeapFrog (AKDK) performs poorly for an adaptive time-step integrator, unless the time-step assigning function is properly selected with a possible computational cost.

%------------------------Figure----------------------------------%

\begin{figure}
    \centering
    \includegraphics[width=0.47\textwidth,angle=0]{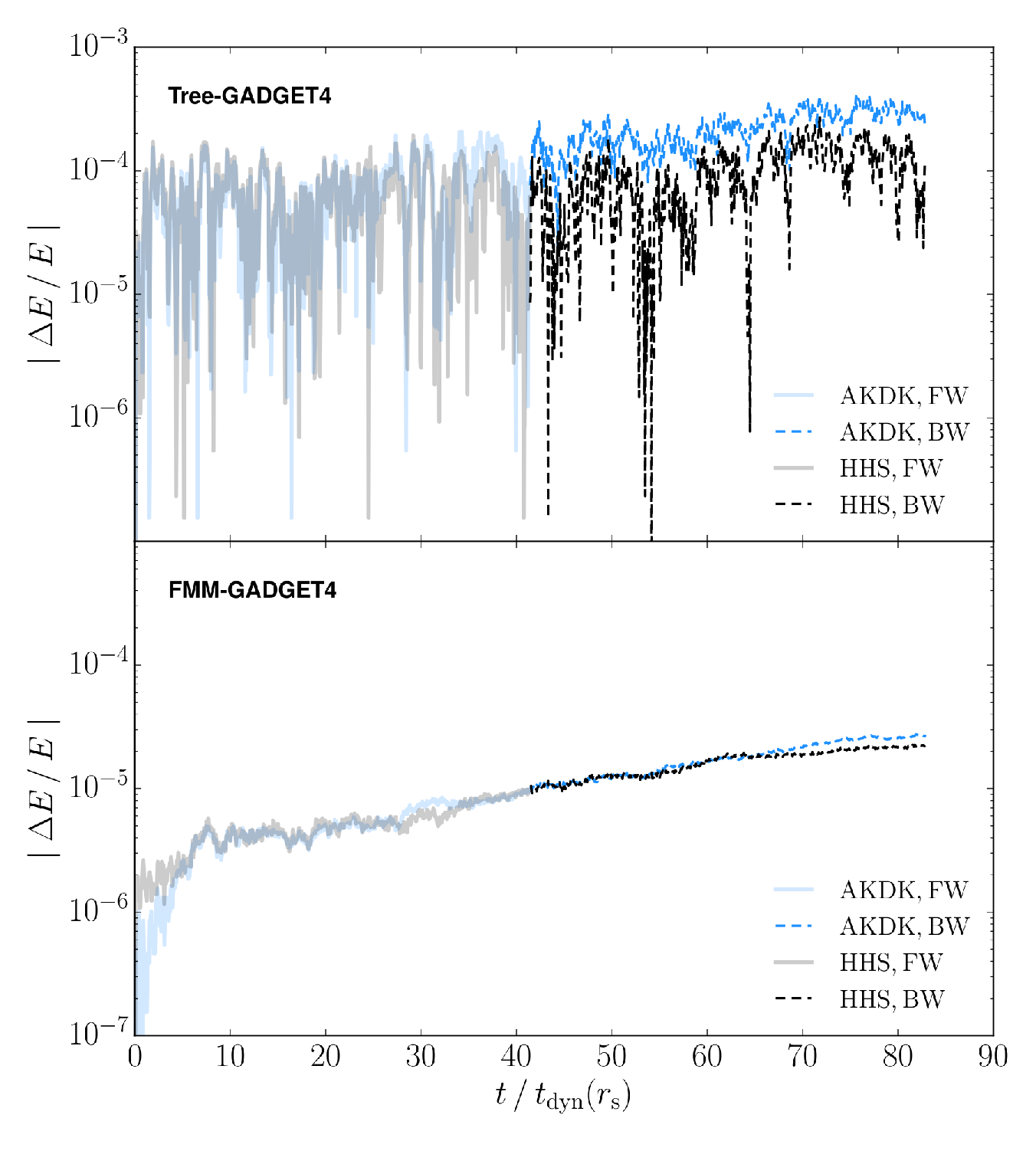}
    \caption{Reversibility test for GADGET4. We followed the fiducial isolated halo using the GADGET4 code (shown with solid opaque lines) and we reversed the sign of particles velocities at around 40 $t_\mathrm{dyn}$ (dashed lines). Upper panel (Tree-GADGET4): for HHS with a non-symmetrized step function, remarkably up to 60 $t_\mathrm{dyn}$ the energy drift stays flat, then it slowly starts to growth. For AKDK the reversed integration just continues the forward systematic growth. Lower panel: For FMM with HHS and AKDK using a non-symmetrized step function, the energy drift grows systematically for both integrators. We can see that Tree-GADGET4 with HHS and the non-symmetrized step function could be approximately reversible or at least more stable, however, more studies is required. }
    \label{fig:rev_G4Tree}
\end{figure}

Figure \ref{fig:ener_sFunc} shows that even GADGET4 with the selection function given by equation \ref{eq:standard} and its implementation of HHS allows, under certain conditions, for a faster or more accurate integration than AKDK. Figure \ref{fig:rev_G4Tree} investigate the reversibility of HHS and AKDK in the Tree (upper panel) and FMM (lower panel) implementations using GADGET4. The forward (FW) evolution of both integrators is shown with solid opaque lines, and the backward (BW) evolution after changing the velocity sign, at 40 $t_{dyn}$, is shown with dashed lines. For the Tree gravity solver using HHS, the energy drift for both forward and backward evolution is similar particularly during the first 10 $t_{dyn}$, however, although encouraging, we do not have enough evidence to claim that HHS is reversible, the time-step function adopted by GADGET4 may be the reason is slightly different to our tests with a direct summation code. On the other hand, the reversed integration of AKDK shows a discontinuous change in its slope, indicating that it is not reversible, we conclude that Tree-HHS is more stable than Tree-AKDK. For the FMM gravity solver the test is not conclusive, because it does not show a significantly smaller slope in energy drift using the HHS scheme (Hierarchical Gravity).

%------------------------Figure----------------------------------%

\section{Discussion and Conclusions} \label{Conclusions}

Using a GPU direct summation \emph{N}-body code, we tested and characterized the Hierarchical Hamiltonian Splitting (HHS) integrator proposed by \citet{2012NewA...17..711P}, but we focused on collisionless simulations. As a reference we compared with the global-constant time-step symplectic Leapfrog integrator and the widely used Adaptive one (AKDK). We also complemented our study using the HHS implementation in GADGET4 \citep[dubbed as hierarchical gravity,][]{2021MNRAS.506.2871S}, which uses different time-step selection function and approximate force solvers (Tree and FMM).

As recently discussed \citep[e.g.][]{2017MNRAS.472.1226D}, there is no general solution for a symplectic adaptive multi time-step integrator, although there are several proposals \citep[e.g.][]{Huang1997,HAIRER1997219,CALVO19981,HARDY199919,2007ApJ...663.1420F}. The problem is not exclusive of astrophysics, the field of differential equations of dynamical systems has extensively reviewed the subject \citep{2002Springer..}. In particular it is important to say that adaptive and multi-step techniques are not identical but they are related. \citet{2018MNRAS.475.5570H} address the important case of adaptive time steps, inspired by \citet{2002Springer..}, they discuss what is called the ``backwards error analysis'' in order to explore commonly used integrators in astrophysics, concluding that symplecticity, time symmetry  and reversibility are not the same and they are not always a guarantee of energy conservation. Nevertheless, KAM (Kolmogorov-Arnold-Moser) theorem assures stability for symplectic (geometrical) and reversible integrators \citep{2002Springer..} in contrast to non-reversible ones, like AKDK with the standard time-step selection function. The case of multi-step integration, requires extra conditions like the so called impulse (splitting) or averaging (less often force evaluation) techniques \citep{2002Springer..}. Similar problems have been discussed in the molecular dynamics field, and the r-RESPA method \citep{1992JChPh..97.1990T}that is an splitting technique has been applied to the long-short range splitting case or multiple mass segregation with considerable advantage in performance. An important feature is that such approaches to the multi-step problem in such cases resulted in a symmetric or reversible method not a symplectic one. The major concrete benefit has been improvement in the performance at a given energy accuracy. A similar situation applies to HHS, the case we study in this paper. HHS is a composition of \emph{Fast} and \emph{Slow} operators \citep{2012NewA...17..711P} and based in our numerical experiments it is approximately reversible, which explains the performance advantage shown in section \ref{Perfomance}. The discussion in \citet{2002Springer..} and \citet{2018MNRAS.475.5570H} shows that the reversibility depends on the symmetry under a velocity sign change of hamiltonian and the corresponding time-step assigning function, and not on the accuracy of gravity calculation or particle number. The particular case of equation \ref{eq:freefall} is an approximated 1st order version of the one proposed by \citet{1995ApJ...443L..93H} that depends only on the module of particle pairs relative velocities, therefore is reversible. A future interesting avenue inspired by the r-RESPA method presented by \citet{1992JChPh..97.1990T} is the extension already discussed by \citet{2020CNSNS..8505240P}. We experimented with the strategy motivated by the multi-step averaging technique \citep{2002Springer..}, updating the time-step hierarchy only at a given number of global time-steps (dubbed as HHS-sTSS, figures \ref{fig:HA_ener_big} and \ref{fig:HA_long}), there is a gain in CPU time, however, the energy drift increases suggesting that further investigation is required. 

Our results are summarized below:

\begin{enumerate}

    \item Based on reversibility and time symmetry tests we concluded that HHS is not time symmetric but it is approximately reversible, it is also more stable than AKDK for a given energy accuracy. Although the exact correspondence between forward and backwards integration lasts only for few dynamical times, even for ten dynamical times the energy drift growth is small. In contrast, AKDK energy drift grows systematically with time, clearly showing that it is non-reversible using the commonly used time-step selection function. Based in our tests HHS reversibility explains the advantage in performance regarding AKDK. Based on \citet{2002Springer..} discussion and the used time-step selection functions that depend only on velocity module, such properties are independent of accuracy in gravity calculation and particle number.
   
    \item Our findings with the direct summation code may be also relevant for the high accuracy and costly section (PP) of codes using the P3M technique, as it has been also show in molecular dynamics studies \citep{Plimpton1997ParticleMeshEA}.
    
    \item In agreement with our direct summation code tests, changing the time-step selection function for a non-symmetrized one, it is possible to find a combination of parameters in GADGET4 (using both the Tree and FMM code version) where HHS is more efficient than AKDK. We found approximated reversibility for the Tree Gadget4 test, however, for FMM Gadget4 our tests are inconclusive. 
    
    \item The population of particle histogram across the time-step rungs is useful to find a convenient parameters combination for the integrators.

\end{enumerate}

\section*{Acknowledgements}
 GAA thanks V. Springel for nicely clarifying GADGET4 information. GAA and JCC thanks Leobardo Ithehua Rico for the fruitful technical support and discussion. OV and GAA acknowledge support from UNAM PAPIIT grant IN112518. GAA and JAT thanks support from CONACyT graduate fellowships 455336 and 825451, respectively. OV acknowledge support from UNAM PAPIIT grants IG101620 and IG101222. HV acknowledges support from IN101918 PAPIIT-UNAM Grant and from a Megaproyecto DGTIC-LANCAD. The authors acknowledge DGTIC-UNAM for access to the Supercomputer MIZTLI. This research was partially supported through computational and human resources provided by the LAMOD UNAM project through the clusters Atocatl and Tochtli. LAMOD is a collaborative effort between the IA, ICN and IQ institutes at UNAM, and it acknowledges benefit from the MX28274 agreement with IBM. The authors acknowledge CEDIA GPU cluster through an agreement with CUDI.

\bibliography{mybibfile}

\end{document}